\documentclass[twocolumn,prd]{revtex4}
\usepackage{graphicx}
\usepackage{dcolumn}
\usepackage{bm}

\begin{document}

\preprint{IFUP-TH/06-18}

\title{ANALYTIC ESTIMATE OF THE ORDER PARAMETER FOR MONOPOLE CONDENSATION IN QCD}

\author{Adriano Di Giacomo}
\email{adriano.digiacomo@df.unipi.it}
\affiliation{ Dipartimento di Fisica Universita di Pisa and INFN Sezione di Pisa ,3 Largo B. Pontecorvo 56127 Pisa (ITALY)
}%

\begin{abstract}{The disorder parameter  $\langle \mu \rangle$  for the condensation of monopoles in QCD is estimated analytically in terms of gauge invariant field strength correlators. The continuum limit is discussed.}
\end{abstract}

\maketitle

\section{Introduction}
It is a long standing conjecture that confinement of color is due to dual superconductivity of  $QDC$ vacuum\cite{1}\cite{2}\cite{3}.

Dual superconductivity means condensation in the vacuum of magnetic charges, and can be detected by use of an order parameter $\langle \mu \rangle$ , which is the vacuum expectation value of a magnetically charged operator $\mu$ \cite{4}\cite{ddp}\cite{ddpp}\cite{dp}. For gauge group $SU(N)$ there are  $N-1$ different magnetic charges and   $\mu^a , (a=1,...,N-1),$  are the corresponding $\mu$ operators.  A gauge invariant definition of  $\mu^a $ can be given and the order parameters  $\langle \mu^a\rangle$
have been studied numerically on the lattice \cite{5} \cite{6}. The independence of  $\langle\mu^a\rangle$ on the abelian projection which defines the monopoles has been argued \cite{7} \cite{71} and demonstrated numerically \cite{8}.

Lattice simulations show that  $\langle \mu^a \rangle\ne 0$ in the confined phase and that  $\langle \mu^a \rangle = 0$ in the deconfined phase where magnetic charge is superselected\cite{9'}, both for quenched \cite{5} \cite{6}  and for unquenched QCD \cite{9}. The scaling behavior of  $\langle \mu^a \rangle$ with respect to volume in the vicinity of the deconfining transition allows to determine the order and the critical indexes of the phase transition \cite{5} \cite{6} \cite{9}.
In this paper I present an analytic computation of $\langle \mu^a \rangle$  in terms of gauge invariant correlators of the gluon field strengths in the vacuum. The aim is to relate confinement to properties of the correlators, which have been extensively studied on the lattice\cite{10} \cite{11} \cite{12}.

The approximation that we shall use in our computation is the same which is at the basis of the "stochastic vacuum" approach to QCD\cite{13a}\cite{13b}\cite{13}: we shall perform a cluster expansion of $\langle \mu^a \rangle$ in terms of the correlators of the field strengths and we shall truncate it at the second cumulant, bilinear in the fields. We expect this approximation to work reasonably in studying infrared properties , which involve correlations at large distances.

In Sect. 2 we recall the definition and the properties of  $\langle \mu^a \rangle$ and we perform its cluster expansion.

In sect. 3  we recollect what is known from lattice simulations on the two-point field correlators, and we compute the susceptibilities $\rho^a$
\begin{equation}
\rho^a = {\partial\over \partial\beta} ln\langle \mu^a \rangle
\end{equation}
which are convenient tools to describe dual superconductivity and the deconfining transition \cite{ddp}\cite{ddpp}. Some computational details are presented in the appendix..

In Sect. 4  we discuss the results.  Confinement and deconfinement are related to properties of the field correlators. We discuss in what respect the numerical determination of the latter can be improved.
As a byproduct we discuss the continuum limit  of  $\rho^a$ and of  $\langle \mu^a \rangle$ and their renormalization.  A similar behavior is known for the Polyakov line in quenched theory \cite{biel}.
\section{Cluster expansion of $\langle \mu^a \rangle$.}
\subsection {\label {A} The order parameter.}

In $SU(N)$ gauge theories , with or without dynamical quarks, the operator which creates a monopole  of magnetic charge  ${q\over {2g}}$  of species  $a$ ,$  (a=1,...,N-1),$ at $(\vec x,t)$ is written as \cite{4}\cite{5}\cite{6}:
\begin{equation}
\mu^a(\vec x, t)= exp[ {q\over{2g}}i\int d^3y \vec b_\perp(\vec x- \vec y )Tr( \Phi^a(\vec y,t)\vec E(\vec y,t)) ]
\end{equation}
$g$ is the coupling constant, $q$  is an integer.  $\vec b_\perp(\vec x - \vec y)$ is the vector potential produced at $\vec y$ by a static monopole sitting at $\vec x$, with
\begin{eqnarray}
\vec \nabla \vec b_\perp& =& 0  \\
\vec \nabla \wedge \vec b_\perp& =&\vec H
\end{eqnarray}
$\vec H$ is the magnetic field, including the Dirac string.

Explicitely, putting the string along the direction $\vec n$ ,
\begin{equation}
\vec b_\perp(\vec x) = { {\vec n \wedge \vec x}\over {x(x - \vec n.\vec x)}}
\end{equation}
$\Phi^a(\vec y,t)$  is a scalar field in the adjoint representation which has the form\cite{16}
\begin{equation}
\Phi^a(\vec y,t)  =  U(\vec y,t)\Phi^a_{diag} U^{\dagger}(\vec y,t)
\end{equation}
with U a generic gauge transformation and $\Phi^a_{diag} $ a diagonal matrix of the form
\begin{eqnarray}
\Phi^a_{diag}=( {N-a\over N}, . . .  ,{N-a\over N}&,&{a\over N}. . . . . . . . . .{a\over N}) \\
                           \leftarrow ---a  --\rightarrow&,&\leftarrow(N-a)\rightarrow \nonumber
\end{eqnarray}
 $U$ can be taken as a parallel transport from a reference point  $x_0$ to $(\vec y,t)$ along a line $C$,
 $U(x) =P exp[ig\int_{x_0,C}^x A_{\mu} dx^{\mu}]$.
 The gauge transformation $U(x)$  identifies the  abelian projection. 
 
 In the gauge in which $\Phi^a(\vec y,t)$ is diagonal , (abelian projected gauge), the electric field which enters in eq(2) is the color component of the transverse electric field along the diagonal generator $T^a$ defined as
 \begin{eqnarray}
 T^a =  diag( 0,0,...,1&,&-1,0,0.....0)\\
            a&,&a+1 \nonumber
  \end{eqnarray}
  since   $Tr(\Phi^a,T^b)  =  \delta^{ab}$.
  
  $\vec E^a_{\perp}$ is nothing but the conjugate momentum to $\vec A^a_{\perp}$ so that $\mu^a$
  simply adds the field of a monopole of charge $q$ to $\vec A^a_{\perp}$.
  \subsection {\label {B} Cluster expansion.}
  From eq(2) 
  \begin{eqnarray}
  \langle \mu \rangle = 1 +  \sum ^{\infty}_{n=1}({iq\over 2g})^n {1\over {n!}}\int d^3 \vec y_1...\int d^3 \vec y_n b^{i_1}_\perp (\vec x-\vec y_1).. \nonumber \\ b^{i_n}_\perp (\vec x- \vec y_n)
\langle (\Phi^a.\vec E)^{i_1}(\vec y_1,t)....(\Phi^a.\vec E)^{i_n}(\vec y_n,t) \rangle
  \end{eqnarray}
  where we have put  $(\Phi^a.\vec E)(\vec y,t)\equiv Tr(\Phi^a . \vec E(\vec y,t))$ in order to simplify the notation.
  
  The term in the expansion eq(9) linear in the electric field is zero by symmetry. In the "gaussian" approximation of the stochastic vacuum model only two point correlators and the even terms survive.
   The combinatorial factor is $(2n-1)!!$ so that
   \begin{eqnarray}
   \langle \mu^a \rangle \simeq exp[ -{q^2\over {8g^2}}\int d^3 y_1..\int d^3y_n b^{i_1}_{\perp}(\vec x-\vec y_1).. \nonumber \\ b^{i_n}_{\perp}(\vec x-\vec y_n)\langle(\Phi^a.\vec E)^{i_1}(\vec y_1,t)..(\Phi^a.\vec E)^{i_2}(\vec y_2,t)\rangle]
   \end{eqnarray}
   or, changing variable to $\vec y_i-\vec x$ , using the translation invariance of the correlator, and the usual definition
   \begin{equation}
   \beta =  {2N\over g^2}
   \end{equation}
   we get 
   \begin{eqnarray}
   \langle \mu^a \rangle \simeq exp[ - {{\beta q^2}\over {16N}}\int d^3y_1\int d^3y_2 b^{i_1}_{\perp}(\vec y_1)b^{i_2}_{\perp}(\vec y_2) \nonumber \\   \Phi^a_{i_1,i_2}(\vec y_1- \vec y_2)]
   \end{eqnarray}
   where  $\Phi^a_{i,j}$ is a gauge invariant correlator which coincides with the correlator of the field strength component along the color space direction  $a$  of eq(8) in the abelian projected gauge.
   
   We shall identify  $\Phi^a_{i,j}(\vec x)$ with the gauge invariant correlator as measured on the lattice \cite{10} \cite{11}\cite{12}. The lattice correlators are defined by connecting the two field strengths by a straight line parallel transport. In general the correlators do depend on the particular path used for parallel transport \cite{megg} . However in what follows we shall only use their qualitative behavior at short and at large distances, which should be independent of it.
   
   In the analysis of confinement it proves convenient to deal not with $\langle \mu^a\rangle $ itself, but with the related susceptibility defined by eq(1) . From eq(12) we get
   \begin{eqnarray}
   \rho^a = - {q^2\over {16N}} {\partial \over {\partial\beta}}[\beta\int \int d^3\vec y_1 d^3\vec y_2 b^{i_1}_{\perp}(\vec y_1)b^{i_2}_{\perp}(\vec y_2) \nonumber \\ \Phi^a_{i_1,i_2}(\vec y_1-\vec y_2)]
   \end{eqnarray}
   
   Eq(13) suggests a quantitative check of the gaussian approximation in the cluster expansion.
   Higher clusters would introduce terms proportional to higher powers of  $q$ . If  $\rho^a$ proves
   to be proportional to $q^2$ to a good approximation higher clusters can be neglected. Old data on $\rho^a$ are consistent  with that, but a careful and systematic numerical check is planned.
  \subsection {\label {C} $\rho^a$ from lattice.} 
   We recall that , since  $\langle \mu^a\rangle  ={ Z(S+\Delta S) \over Z(S)},  $ $\langle \mu^a\rangle =1$ at $\beta =0$ and therefore
   \begin{equation}
   \langle \mu^a\rangle = exp[ \int^{\beta}_0 \rho(\beta') d{\beta'}]
   \end{equation}
   Lattice determinations of $\rho^a$ \cite{5}\cite{6}\cite{7}\cite{8} can be summarized as follows. If $\beta_c$ is the critical value of $\beta$ corresponding to the deconfinement transition, in the thermodynamical limit , when the spatial volume $L_s^3$ goes to infinity:
   
   1) For $\beta \le \beta_c$ $\rho^a$ tends to a finite limit , implying by use of eq(14) that $\langle \mu^a\rangle  \neq 0$.
   
   2) For $\beta \ge \beta_c$ 
    \begin{eqnarray}
    \rho^a\simeq c' - c L_s  \\
    c    >     0   \nonumber
    \end{eqnarray}
    implying, again by use of eq(14) that $\langle \mu^a\rangle=0$ in the thermodynamical limit.
    
    3) For  $\beta \cong  \beta_c$  $\rho^a$ obeys the scaling law \cite{5}\cite{6}\cite{17}
    \begin{equation}
    {\rho^a\over L_s^{1\over \nu}} = f( \tau  L_s^{1\over \nu})
    \end{equation}
    with $\tau \equiv (1-{T\over T_c})$ the reduced temperature and $\nu$ the critical exponent of the
    correlation length , $\lambda$ , of the order parameter.
   \begin{equation}
    \lambda   { \approx }_{\tau \to 0} \,\,\,    \tau^{-\nu}
    \end{equation}
    We shall discuss how these features can emerge from eq(13).
   
\section{Computing $\rho^a$.}

By general invariance arguments the electric field correlator at equal times has the form [ see e.g. eq(2.6) of ref . 13]
\begin{equation}
{1\over N} \Phi^a_{ij}(\vec z)  =  \delta_{ij} [D(z^2) + {1\over2} D_1(z^2) ]  + \partial_i[z_j.D_1(z^2)]
\end{equation}
with $D,D_1$ two invariant functions of $k^2$.
The dependence on $a, (a=1,..N-1)$ disappears in eq(12) : this is in agreement with what is observed numerically\cite{5}\cite{6}\cite{17}. Notice that this is true if we identify our gauge invariant correlators with
the ones of the stochastic model as discussed above. A dependence on  $a$ ,however, could in principle exist  in our correlators.
The last term in eq(18) does not contribute to the convolution with $\vec b_{\perp}$ in eq(13) 
since $\vec \nabla.\vec b_{\perp} = 0$. After Fourier transform eq(13) becomes then
\begin{equation}
\rho^a = - {q^2\over16} {\partial\over {\partial\beta}}[\beta \int{d^3k\over (2\pi)^3} \tilde b^i_{\perp}(\vec k)\tilde b^j_{\perp}(-\vec k)\tilde\Phi^a_{ij}(k^2) ]
\end{equation}
with
\begin{equation}
\Phi^a_{ij}(k^2) = \delta_{ij}[ \tilde D(k^2) + {1\over2}\tilde D_1(k^2)]
\end{equation}
Since $\vec k.\vec b_{\perp}(\pm \vec k)=0$ we can replace $\delta_{ij}$ in eq(20) by $k^2\delta_{ij} -{k_ik_j}\over k^2$ and, putting
\begin{equation}
k^2 f(k^2) \equiv \tilde D(k^2) + {1\over 2}\tilde D_1(k^2)
\end{equation}
and making use of the equality
\begin{equation}
(k^2\delta_{ij}  - {k_ik_j})\tilde b^i_{\perp}\tilde b^j_{\perp} = \vec H(\vec k).\vec H(-\vec k)
\end{equation}
with $\vec H(\vec k)$ the Fourier transform of the magnetic field, we finally obtain
\begin{equation}
\rho^a = - {q^2\over 16}{\partial\over {\partial\beta}}[\beta \int{d^3k\over {(2\pi)^3}}|\vec H(\vec k)|^2 f(k^2)]
\end{equation}
If the Dirac string of the monopole is put along the direction $\vec n$ , say $z$,then
\begin{equation}
\vec b_{\perp}(\vec k) = {{\vec n\wedge \vec k}\over{k(k-\vec k.\vec n -i \epsilon)}}
\end{equation}
and
\begin{equation}
\vec H(\vec k)= -i\vec k \wedge \vec b_{\perp}(\vec k)= -i[{\vec k\over k^2} - {\vec n\over {(\vec n.\vec k)-i\epsilon}}]
\end{equation}
or
\begin{equation}
|\vec H(\vec k)|^2 = {1\over k_z^2}  - {1\over k^2}
\end{equation}
and
\begin{equation}
\rho = {q^2\over 16N}{\partial\over {\partial\beta}}[\beta\int {d^3k\over {(2\pi)^3}} f(k^2)({1\over k^2} - {1\over k_z^2})]
\end{equation}
At large $\beta$ (high temperature) $f(k^2)$ can be approximated by lowest order perturbation theory
which is easily computed to be $ f(k^2) = {1\over {2k}}$ , the dependence on $\beta$ in eq(25) only comes from the explicit factor $\beta$ and
\begin{equation}
\rho = {q^2\over{16N}} \int{d^3k\over {(2\pi)^3}}{1\over{2k}}({1\over k^2} - {1\over k_z^2})
\end{equation}
The integral is easily  computed [See Appendix A] with an ultraviolet cut-off   $ {1\over a} $    and an infrared cut-off    ${1\over {L_s.a}}$    giving
\begin{equation}
\rho = {q^2\over{16N}}{1\over {(2\pi)^2}}[ -\sqrt2 L_s + 2 lnL_s + constant]
\end{equation}
Eq(29) means $\langle \mu\rangle = 0$ as $L_s\to \infty$ , or superselection of magnetic charge.
This is indeed what is observed in lattice simulations \cite{5}\cite{6}\cite{9}\cite{9'}.
Also the proportionality  to $q^2$ of the slope  with respect to $L_s$  is consistent with data\cite{9'}.

The field correlators  $D$ and $D_1$ as measured on the lattice in the range of distances $.1fm\leftrightarrow1fm$ are well fitted by a parametrization of the form\cite{10}\cite{18}:
\begin{eqnarray}
D & =&  A_0exp(-{x\over \lambda_b})  + {b_0\over {x^4}}exp(-{x\over \lambda_a})\\
D_1&=& A_1exp(-{x\over \lambda_b})  + {b_1\over {x^4}}exp(-{x\over \lambda_a})
\end{eqnarray}
The slopes of both exponentials in $D$ and $D_1$ are equal within errors: $\lambda_b\approx .3fm$,
$\lambda_a\approx 2\lambda_b$.

The parametrization eqs(30)-(31) is inspired by the Operator Product Expansion (OPE). The terms proportional to  ${1\over{x^4}}$ correspond to the identity operator, the terms with only the exponential
to the gluon condensate $G_2$ : more precisely  $ A + A_1 ={\pi^2\over18} G_2$ with $G_2 =\langle {\beta(g)\over g} G^a_{\mu\nu}G^a_{\mu\nu}\rangle$.

In fact the parametrization eqs(30)-(31) cannot be correct at short distances since the expansion of the second terms gives non-zero coefficients to terms  in  ${1\over {x^3}},{1\over {x^2}},{1\over x}$ implying the existence of condensates of dimension $1,2,3,$, which do not exist. Similarly for the exponential terms the expansion of the exponential implies the existence of a condensate of dimension $5$ ,which again does not exist. The correct OPE should have the form 
\begin{equation}
D (D_1) \approx_{x\to 0} {b\over {x^4}} + A + Cx^2+..
\end{equation}
or better, to match the perturbative expansion at small distances
\begin{equation}
D (D_1) \approx_{x\to 0} {b\over 2}[{1\over {(x+i\epsilon )^4}}+{1\over {(x-i\epsilon )^4}}] + A + Cx^2+..
\end{equation}
A possible simple form could be obtained by the replacements
\begin{eqnarray}
\exp-({x\over \lambda_a})   \longrightarrow      \exp-([({x\over \lambda_a})^4 +\epsilon]^{1\over 4}-\epsilon)\\
\exp-({x\over \lambda_b})    \longrightarrow       \exp-([({x\over \lambda_b})^2 +\epsilon_1]^{1\over 2}-\epsilon_1)
\end{eqnarray}
with suitable values  of $\epsilon$, $\epsilon'$ to match lattice data within errors.
 At large distances a strong infrared cut-off must exist in the confined phase , which should disappear at the deconfining transition. Such a cut-off implies for the term in $A +{A_1\over 2}$ that the space integral must be zero.A change of the form
\begin{equation}
A exp(-{x\over{\lambda_b}})  \longrightarrow  A(1 - {x^2\over{12\lambda^2_b}}) exp(-{x\over{\lambda_b}})
\end{equation}
is  consistent with the lattice data on correlators within errors, and give zero integral. Parametrizations of the correlators which imply a change of sign at large distances exist in the literature\cite{dosch'}.

 Since the short distance behavior coincides with the perturbative one  the singular term will give a result similar to that of eq(29), with $L_s$ replaced by the infrared cut-off $\Lambda$ and a different value for the constant. The contribution of  exponential term modified as in eq(34) is easily computed . The total result is
\begin{eqnarray}
\rho = {q^2\over {16N}} {\partial\over {\partial\beta}}{\beta}[ {1\over{(2\pi)^2}}( -\sqrt2{\Lambda \over a} +2ln({\Lambda \over a}) + const.) \nonumber \\-{7\over {3\pi}}
N{\lambda_b}^3{\Lambda} (A+{1\over 2}A_1)]
\end{eqnarray}
From zero temperature up to say $.95T_c$ the correlators are constant,
the only dependence on $\beta$ is the explicit one in eq(27) so that
\begin{eqnarray}
\rho = {q^2\over {16N}}[ {1\over{(2\pi)^2}}( -\sqrt2{\Lambda \over a} +2ln({\Lambda \over a}) + const) \nonumber \\-{7\over {3\pi}}
N{\lambda_b}^3{\Lambda} (A+{1\over 2}A_1)]
\end{eqnarray}
 which is finite and volume independent, so that  , by use of eq(14)  $\langle \mu \rangle \neq 0$ (dual superconductivity).
 
 This is true at fixed ultraviolet cut-off. In the continuum limit, however, when $a \to 0$ $\rho \propto {1\over a}$ and a renormalization is needed like for the Polyakov line \cite{biel}. Existing  lattice data are consistent with this behavior  ( see e.g. fig .2    of ref[8] ), but an extensive numerical check is planned.
 
  As $T \to T_c$  $\Lambda \to \infty$ , the electric condensate $\langle \vec E^2\rangle$ , and hence $A$ and $A_1$ vanish very rapidly \cite{13}   [See e.g. Fig . 10 of ref [26].] $\lambda_a, \lambda_b$ do not change appreciably across $T_c$. $\Lambda $ diverges when approaching the critical point $\tau = 0$. Eq(37)
  must then be used.  As a model example if
  \begin{eqnarray}
  {\Lambda\over a} &&  \approx    K ln(\tau)   \\
     && \tau \to 0  \nonumber
   \end{eqnarray}
   the singular part only comes from the first term in parenthesis of eq(37) giving
   \begin{equation}
   \rho = - {{\sqrt2 Kq^2}\over {16N(2 \pi)^2}}{{\beta}\over \tau}
   \end{equation}
   This is consistent with scaling eq(16). A quantitative numerical analysis is needed, however,  to have a precise picture of the transition.
   \section{Concluding remarks}
   We have related the susceptibilities $\rho^a$ of the order parameter for dual superconductivity to properties of the gauge invariant field strength correlators, by use of the stochastic vacuum model of QCD, i.e. by a cluster expansion of field strength correlators truncated at $n=2$. A direct check of this approximation can be done by verifying that  $\rho^a$ is proportional to the square of the magnetic charge.
       
    The validity of the perturbative approximation at large temperatures insures superselection of magnetic charge, i.e. that vacuum is normal.  
   
   Dual superconductivity   in the confined phase is related to an infrared cut-off of the two point correlator, which renders   $\rho^a$  volume independent.
   
   The order parameter is anyhow ultraviolet divergent and needs a renormalization in the continuum limit, similarly to what happens to the Polyakov line in the quenched theory.
   
   The scaling behavior in the critical region depends on how the infrared cut-off diverges , and on how the electric condensate vanishes at the critical point.
   A direct monitoring of the large distance behavior of the two-point correlator is needed to confirm
   our assumptions, which go beyond the existing lattice data on correlators, as well as a more precise study of their behavior at the critical point.
   
 \begin{acknowledgments}
I wish to thank M. D'Elia,H. G. Dosch,  E. Meggiolaro and G. Paffuti for useful discussions. The idea of this computation originated from a discussion with Yu. A. Simonov  a couple of years ago, for which I thank him. This work is partially supported by MIUR project  2004 Theory and phenomenology of fundamental interactions.
\end{acknowledgments}

   \appendix
  \section{A}
   We  compute the integral in eq(28) of the text.
   \begin{equation}
I  \equiv  \int{d^3k\over {(2\pi)^3}}{1\over{2k}}({1\over k^2} - {1\over k_z^2})
\end{equation}
We integrate the first term in spherical coordinates, the second one in cylindrical coordinates.
\begin{equation} 
I  = [ \int^{1\over a}_{1\over{L_s a}}{dk\over k}  -\int{dk_z\over k_z^2} {{dk_{\perp}^2}\over {2\sqrt( k_z^2+k_{\perp}^2)}}]
\end{equation}
We have introduced as ultraviolet cut-off on each component of $\vec k$ the inverse of the lattice spacing $a$ and as infrared cut-off the inverse of the spatial lattice edge $L_s a$.
Performing first the integral over $k_{\perp}^2$ from $k_{\perp}^2{min} =  {2\over{a L_s}^2}$ to  $k_{\perp}^2{max} = 2{1\over{a }^2}$ gives  for the second integral in parenthesis 
\begin{equation}
\int ^{{k_z}max}_{{k_z}min}{dk_z\over k_z^2}\sqrt(k_z^2+k_{\perp}^2)|^{k_{\perp}^2{max}}_{k_{\perp}^2{min}}
\end{equation}
Here again ${k_z}max= {1\over a} , {k_zmin}={1\over{a L_s}}$.
Using the indefinite integral 
\begin{equation}
\int {dx\over x^2}\sqrt(x^2 + c^2) = -{\sqrt(x^2 + c^2)\over x} + ln(x+\sqrt(x^2 + c^2))
\end{equation}
one finally gets
\begin{equation}
I = {1\over (2\pi)^2}[ -\sqrt2 L_s +2ln(L_s) +2\sqrt3 - 1 -ln{{(1+\sqrt3)^2}\over {2\sqrt2}}]
\end{equation}
\end{document}